\documentclass[review,3p,autheryear,showpacs,number,report]{elsarticle}

\usepackage[switch,displaymath,mathlines]{lineno}

\def\aeq{&=&}

\usepackage{amsmath,amssymb,amsfonts}
\usepackage{bm, slashed}
\usepackage{color,graphicx}
\usepackage{subfigure}
\usepackage[colorlinks=true]{hyperref}
\usepackage{xcolor}
\definecolor{red}{rgb}{0.9, 0,0}

\usepackage{hyperref}

\newcommand{\bw}{\begin{widetext}}
\newcommand{\ew}{\end{widetext}}
\newcommand{\be}{\begin{equation}}
\newcommand{\en}{\end{equation}}
\newcommand{\bee}{\begin{equation}}
\newcommand{\ene}{\end{equation}}
\newcommand{\bea}{\begin{eqnarray}}
\newcommand{\ena}{\end{eqnarray}}
\newcommand{\eq}[1]{Eq.~(\ref{#1})}

\def\pslash{p\!\!\!\slash }

\def\Sslash{S\!\!\!\slash }

\def\nn{\noindent}
\def\to{\rightarrow}

\def\gev{{\rm GeV}}
\def\pslash{p\!\!\!\slash }

\begin{document}

\title{Seiberg-Witten map and quantum phase effects for neutral Dirac particle on noncommutatiave plane}

\author{Kai Ma}
\ead{makainca@yeah.net}

\author{Jian-Hua Wang}
\address{Department of Physics, Shaanxi University  of Technology, Hanzhong, 723001, Peoples Republic of China}

\author{Huan-Xiong Yang}
\address{Interdisciplinary Center for Theoretical Study, University of Science and Technology of China, Hefei 200026, Peoples Republic of China}

\date{\today}

\begin{abstract}

We provide a new approach to study the noncommutative effects on the neutral Dirac particle with anomalous magnetic or electric dipole moment on the noncommutative plane. The advantages of this approach are demonstrated by investigating the noncommutative corrections on the Aharonov-Casher and He-McKellar-Wilkens effects. This approach is based on the effective $U(1)$ gauge symmetry for the electrodynamics of spin on the two dimensional space. The Seiberg-Witten map for this symmetry is then employed when we study the noncommutative corrections. Because the Seiberg-Witten map preserves the gauge symmetry, the noncommutative corrections can be defined consistently with the ordinary phases. Based on this approach we find the noncommutative corrections on the Aharonov-Casher and He-McKellar-Wilkens phases consist of two terms. The first one depends on the beam particle velocity and consistence with the previous results. However the second term is velocity-independent and then completely new. Therefore our results indicate it is possible to investigate the noncommutative space by using ultra-cold neutron interferometer in which the velocity-dependent term is negligible. Furthermore, both these two terms are proportional to the ratio between the noncommutative parameter $\theta$ and the cross section $A_{e/m}$ of the electrical/magnetic charged line enclosed by the trajectory of beam particles. Therefore the experimental sensitivity can be significantly enhanced by reduce the cross section of the charge line $A_{e/m}$. 

\end{abstract}


\maketitle
\tableofcontents
\setcounter{page}{1}
\renewcommand{\thefootnote}{\arabic{footnote}}
\setcounter{footnote}{1}

\section{Introduction}\label{intro}

The noncommutative geometry has been realized as one of the most possible physical structure of the background space-time. The noncommutativity of space can be realized in the dynamics of charged particles in the presence of electromagnetic fields \cite{Peierls:1933}. Snyder pointed out that Lorentz invariant noncommutative space-time can be constructed by using coordinates with discrete spectrum \cite{Snyder:1947}. The noncommutative space-time was also motivated by the string theory \cite{SW:1999} and quantum gravity \cite{Freidel:2006}. 
The noncommutativity of space-time is characterized by the commutation relation $ [x_{\mu}, x_{\nu}] = i \theta_{\mu\nu}$, with the totally anti-symmetric constant tensor parameter $\theta_{\mu\nu}$ representing the strength and relative directions of the noncommutativity\cite{Douglas:2001, Szabo:2003}. As a consequence, the ordinary product is deformed and replaced by the $\star$-product,

\begin{equation}\label{eq:starproduct}
f( x ) \star g( x )
= \exp{\left[\frac{i}{2} \theta_{\mu\nu} \partial_{x_{\mu}}\partial_{y_{\nu}}\right]} f(x)g(y)|_{x=y},
\end{equation}

\nn where $f(x)$ and $g(x)$ are two arbitrary infinitely differentiable functions on the commutative $R^{3+1}$ space-time. Based on the $\star$-product, the gauge theory can be established in the usual way. The $\star$-product can also be represented by the transformation $x^{\mu} \to x^{\mu} + \theta^{\mu\nu}p_{\nu}/2\hbar$, which is called the Bopp's shift\cite{TCurtright:1998}. 

In last decade, the phenomenology of the noncommutative space have been investigated extensively \cite{Douglas:2001}-\cite{YHXang:2008}. Particluarly, it was pointed out that the topological Aharonov-Bohm \cite{AB:1959}  and Aharonov-Casher \cite{AC:1984} effects are sensitive to the noncommutative corrections. In Refs.~\cite{Chaichian:2002, Chaichian:2001B, Li:2006}, the authors  studied the Aharonov-Bohm (AB) \cite{AB:1959} phase on a noncommutative space. A lower bound $1/\sqrt\theta \geq 10^{-6}\gev$ for the space  noncommutativity parameter is obtained. The Aharonov-Casher (AC) \cite{AC:1984} phases for spin-1/2 and spin-1 particles on both noncommutative space (NCS) and noncommutative phase space (NCPS) have also been studied in Refs.~\cite{Mirza:2004, Li:2007,Mirza:2006}. In addition, the noncommutative corrections on the He-McKellar-Wilkens (HMW) \cite{HM:1991} phase is studied in Refs.\cite{Wang:2007,Li:2008}. However, the previous studies based on the Bopp's shift method shown that the noncommutative corrections on the AC and HMW phases are proportional to the beam velocity of the neutron interferometer. Therefore the experimental sensitivity is significantly reduced in the ultra-cold neutron interferometer which has best precision because of the better beam coherence and interference intensity\cite{DirkDubbers:2011,ADCronin:2009}. 

In this paper, in stead of using the Bopp's shift, we employ the Seiberg-Witten (SW) \cite{SW:1999} map from the noncommmutative space-time to the ordinary one, which has been proved to be very useful to investigate various problems on noncommutative space \cite{Martin:2012aw, Brace:2001, Barnich:2001, Picariello:2002, Banerjee:2002,Fidanza:2002,Jackiw:2002}, to study the noncommutative corrections on the AC and HMW phase effects. We will shown that there is an additional correction term that is velocity-independent. We use the SW map because it preserves the ordinary gauge symmetry, which is essential for defining the topological AC and HMW phases\cite{HM:1991,HM:2001,KaiMa:2012,DouglasSingleton:2015}. Here we first review the AC and HMW effects in ordinary $2+1$ dimensions in section \ref{ACHMW}, and interpret these effects as consequences of an effective $U(1)$ gauge symmetry. Based on this result, we introduce the SW map which preserves this effective gauge symmetry. Then the physical observables for the AC and HMW effects on noncommutative space can be defined uniquely, and related to the original phases unambiguously. This parts are studied in section \ref{NCACHMW}. Our conclusions and discussions are given in the final section \ref{conclusion}.

\section{Effective $U(1)$ symmetry for neutral spinor in $2+1$ dimensions}\label{ACHMW}

\subsection{Effective $U(1)$ symmetry}

In this section we review the AC and HMW effects in the $2+1$ dimensions which has been studied seriously by He and Mckellar \cite{HM:2001}. In their paper, the AC and HMW effects were interpreted as the interaction between electromagnetic fields $F_{\mu\nu}$ and dual current $\epsilon^{\mu\nu\alpha}j_{\alpha}$. However, it can also be interpreted as an effective $U(1)$ gauge symmetry with effective gauge potential $S^{\alpha} \propto \epsilon^{\alpha\mu\nu}F_{\mu\nu}$ coupling to the current $j_{\alpha}$. In this paper we first introduce this effective gauge symmetry and interpret the AC and HMW phases as results of this symmetry. Based on this effective $U(1)$ gauge symmetry, we will introduce in next section the corresponding SW map which preserve this effective symmetry and then physical meaning of the observables can be uniquely demonstrated on the noncommutative plane. Let us start with the electromagnetic dynamics of magnetic dipole moment on the ordinary space-time. For convenience, we will call the electromagnetic $U(1)$ gauge symmetry for charged particle as $U_{cha}(1)$ symmetry, and the effective $U(1)$ gauge symmetry introduced here for neutral particle as $U_{neu}(1)$. The Lagrangian for a neutral particle of spin $1/2$ with an anomalous magnetic dipole moment $\mu_{m}$ interacting with the electromagnetic field is given by

\bee\label{lag}
\mathcal{L} =
\bar{\psi}(x)( \pslash - m ) \psi(x) - \frac{1}{2} \mu_{m} \bar{\psi}(x) \sigma^{\mu\nu}\psi(x) F_{\mu\nu}.
\ene

\nn The last term is responsible for the AC effect. The $2+1$ dimensional metric and anti-symmetric tensors are defined as follows,

\bee
g_{\mu\nu} = \rm{diag}(1, -1, -1), ~ \rm{and}~~
\epsilon_{012} = 1.
\ene

\nn and the electromagnetic field strength tensor is

\bee\label{f}
F^{\mu\nu} = 
\left(\begin{array}{ccc}
0 & -E^{1} & -E^{2} \\
E^{1} & 0 & -B^{3} \\
E^{2} & B^{3} & 0 
\end{array}\right),
\ene

\nn where $E^{i}$ and $B^{i}$ are the electric and magnetic fields respectively. The indices ``1" and ``2" indicate the $x$ and $y$ directions on the $x-y$ plane, and the index ``3" indicate the $z$ direction which is normal to the $x-y$ plane. On the $2+1$ dimensional spacetime, there are two inequivalent representations of the Dirac algebra that generate different Clifford algebras \cite{HM:2001}. A particular representation is

\bee\label{rep}
\gamma^{0} = 
\left(\begin{array}{ccc}
\sigma^{3} & 0 \\
0 & \sigma^{3} 
\end{array}\right),~
\gamma^{1} = 
\left(\begin{array}{ccc}
i\sigma^{2} & 0 \\
0 & -i\sigma^{2} 
\end{array}\right),~
\gamma^{2} = 
\left(\begin{array}{ccc}
i\sigma^{1} & 0 \\
0 & i\sigma^{1} 
\end{array}\right).
\ene

\nn In this representation all the $\gamma$-matrix are block diagonal. Furthermore, in the $2+1$ dimensions these Dirac matrices satisfy the following relation

\bee\label{relation}
\gamma_{\mu}\gamma_{\nu} = g_{\mu\nu} + i s \epsilon_{\mu\nu\alpha} \gamma^{\alpha},
\ene

\nn where $s = -i \gamma^{0}\gamma^{1}\gamma^{2} = -\gamma^{0}\sigma^{12}$, which has the eigenvalues $\hat{s} = \pm 1$. In the representation defined by \eqref{rep}, the operator $s$ is diagonal, $s={\rm diag}(I, -I)$. Furthermore, with the relation \eqref{relation}, the $\sigma_{\mu\nu}$ can be simplified $\sigma_{\mu\nu} = - s \epsilon_{\mu\nu\alpha} \gamma^{\alpha}$ which is again diagonal. Then all $\gamma$ matrices involved in the Lagrangian \eqref{lag} are diagonal. This nontrivial feature means that the ``up" and ``down" components of the 4-spinor don't mix with each other, their dynamical evolutions are separated completely. The states are represented by the eigenvalues $\hat{s}$ of the matrix $s$. By using this feature the Lagrangian \eqref{lag} can be written as

\bee
\mathcal{L} = 
\bar{\psi}_{\hat{s}}(x)( \pslash - m ) \psi_{\hat{s}}(x) +
\frac{1}{2} \hat{s} \mu_{m}  \bar{\psi}_{\hat{s}}(x) \gamma_{\alpha}\psi_{\hat{s}}(x) \epsilon^{\alpha\mu\nu} F_{\mu\nu}(x)~,
\ene

\nn here the spinor $\psi_{\hat{s}}$ are eigenstates of the operator $s$ with eigenvalues $\hat{s}$, and we have suppressed the summation over the two different eigenstates. In the following contents, we also suppress the index $\hat{s}$, because all the results are exactly the same except for the eigenvalues $\hat{s}$. By introducing an effective vector potential 

\bee\label{s}
S^{\alpha}(x) = - \frac{1}{2} \epsilon^{\alpha\mu\nu} F_{\mu\nu}(x)~.
\ene

\nn The Lagrangian \eqref{lag} can be rewritten into a minimal-coupling form as follows

\bee\label{miniflag}
\mathcal{L} = \bar{\psi}(x) \big( i \gamma_{\alpha} \mathcal{D}^{\alpha} - m \big)  \psi(x)  ~,~~
 \mathcal{D}^{\alpha} =  \partial^{\alpha} + i  \hat{s} \mu_{m} S^{\alpha},
\ene

\nn where $ \mathcal{D}^{\alpha}$ has been used to denote the effective covariant derivative. Similar to the $U_{cha}(1)$ gauge symmetry of electrodynamics for charged particles, this minimal coupling form means a new $U(1)$ symmetry with the gauge potential $S^{\mu}$, and  $\hat{s}$ is the sign of the ``charge" of matter particle, $\mu_{m}$ is the magnitude of the ``charge". Clearly, the Lagrangian \eqref{lag} is invariant under following gauge transformations,

\bea
\label{effectuo1}
\psi(x) &\to & e^{ -i \mu_{m} \hat{s} \lambda(x) } \psi(x) ~,
\\
\label{effectuo2}
S_{\mu}(x) &\to & S_{\mu}(x) -  \partial_{\mu}\lambda(x) ~.
\ena

Similarly, an effective $U(1)$ gauge symmetry can be introduced in the same way for the electrodynamics of electric dipole moment. The Lagrangian for a neutral particle of spin $1/2$ with an anomalous electric dipole moment $\mu_{e}$ interacting with the electromagnetic field is given by
\bea\label{lagHMW}
\mathcal{L} &=&
\bar{\psi}(x)( \pslash - m ) \psi(x) + \frac{1}{2} \mu_{e} \bar{\psi}(x) \sigma^{\mu\nu}\psi(x) \tilde{F}_{\mu\nu}.
\ena
where $\widetilde{F}_{\mu\nu}= \epsilon_{\mu\nu\alpha\beta}F^{\alpha\beta}/2$ is the $3+1$ dimensional dual of the electromagnetic field tensor. On the $2+1$ dimensional space-time, it has the following expression,
\bea\label{dualf}
\widetilde{F}^{\mu\nu} &=& 
\left(\begin{array}{ccc}
0 & -B^{1} & -B^{2} \\
B^{1} & 0 & E^{3} \\
B^{2} & -E^{3} & 0 
\end{array}\right),
\ena
By using relation \eqref{relation}, and introducing the effective dual gauge potential 
\bea\label{duals}
\widetilde{S} &=&  - \frac{1}{2} \epsilon^{\alpha\mu\nu}  \widetilde{F}_{\mu\nu}(x).
\ena
Then the Lagrangian \eqref{lagHMW} can be written as
\bee\label{miniflag}
\mathcal{L} = \bar{\psi}(x)  ( i \gamma_{\alpha} \tilde{\mathcal{D}}^{\alpha} - m ) \psi(x) ~,~~
\widetilde{\mathcal{D}}^{\alpha} = \partial^{\alpha} + i \hat{s} \mu_{e} \widetilde{S}^{\alpha}(x )~,
\ene
where $ \widetilde{\mathcal{D}}^{\alpha}$ has been used to denote the dual effective covariant derivative. In this case, the gauge potential is $\widetilde{S}^{\mu}$, and  $\hat{s}$ is the sign of the ``charge" of matter particle, $\mu_{e}$ is the magnitude of the ``charge". Clearly, the Lagrangian \eqref{lag} is invariant under following gauge transformations,
\bea
\label{effectuo1}
\psi(x) &\to & e^{ -i \mu_{e} \hat{s} \lambda(x) } \psi(x) ~,  \\
\label{effectuo2}
\widetilde{S}_{\mu}(x) &\to & \widetilde{S}_{\mu}(x) - \partial_{\mu}\lambda(x) ~.
\ena

In summary, we have introduced a new effective $U_{neu}(1)$ gauge symmetries in the electrodynamics of neutral particles with magnetic or electric dipole moment. The interactions are written into the minimal coupling form. We will show in next subsection, the AC and HMW phases can be directly obtained in our approach.

\subsection{Aharonov-Casher and He-Mckellar-Wilkens effects}
Like the AB effect in the ordinary $U_{cha}(1)$ gauge invariant electrodynamics, the neutral particle can also acquire a nontrivial phase factor because of this effective $U_{neu}(1)$ gauge symmetry. From the Lagrangian \eqref{miniflag} we can obtain the equation of motion for the neutral particle with an anomalous magnetic dipole moment,
\bea\label{memm}
( i\gamma_{\mu}\mathcal{D}^{\mu} - m ) \psi(x) \aeq 0
\ena
If $\psi(x)$ is a solution of the free equation, then 
\bea
\psi'(x) \aeq \exp\bigg\{ - i \hat{s} \mu_{m} \int S^{\alpha} dx_{\alpha}  \bigg\}
\ena
is a solution of the equation \eqref{memm}. After a cycler evolution the particle acquires a nontrivial quantum phase 
\bea
\phi \aeq \hat{s} \mu_{m} \oint d x_{\alpha}S^{\alpha} = \frac{1}{2}\hat{s} \mu_{m} \oint d \Omega_{\alpha\beta}W^{\alpha\beta},
\ena
where $d \Omega_{\alpha\beta}$ is the surface enclosed by the orbital of the matter particle, and $W^{\alpha\beta} = \partial^{\alpha}S^{\beta} -  \partial^{\beta}S^{\alpha}$ is the effective field strength. Inserting the electromagnetic fields \eqref{f} into \eqref{s} we get
\bea
S^{\alpha}(x) \aeq ( B^{3}, E^{2}, -E^{1} )~.
\ena
In the AC configuration, the magnetic field $B^{3}$ vanishes and the electric fields $E^{i}$ are constant in time. Then $S^{\alpha}(x) = ( 0, E^{2}, -E^{1} )$, and the corresponding nontrivial AC phase is
\bea
\phi^{AC} 
\aeq - \hat{s} \mu_{m} \oint d \vec{r} \cdot \vec{S} 
= - \hat{s} \mu_{m} \oint d \vec{\Omega} \cdot (\vec{\nabla}\times \vec{S} ) 
= - \hat{s} \mu_{m} \oint d \vec{\Omega} \cdot \vec{W}_{B}
= \hat{s} \mu_{m} \lambda_{e},
\ena
where $\vec{W}_{B}$ is the magnetic component of the field strength, and $\lambda_{e}$ is the charged line density in the setup of AC effect. We can see that the AC phase can be obtained directly in our approach. The topological and nonlocal property of this phase effect can be understood by the dependence on $\lambda_{e}$ which is zero at the location of matter particle. This is the sole property of the AC effect. When we extend it on the noncommutative space, this property has to be preserved in order to uniquely interpret the physical meaning of the phase. We will study this in next section. Let us consider the HMW phase effect for now.

The HMW phase can be obtained in the same way. The equation of motion of the neutral particle with an anomalous electric dipole moment is
\bea\label{medm}
( i\gamma_{\mu}\widetilde{\mathcal{D}}^{\mu} - m ) \psi(x) \aeq 0~.
\ena
If $\psi(x)$ is a solution of the free equation, then 
\bea
\psi'(x) \aeq \exp\bigg\{ - i \hat{s} \mu_{e} \int \widetilde{S}^{\alpha} dx_{\alpha}  \bigg\}
\ena
is a solution of the equation \eqref{medm}. After a cycler evolution the particle acquires a nontrivial quantum phase 
\bea
\phi \aeq - \hat{s} \mu_{e} \oint d x_{\alpha}\widetilde{S}^{\alpha} = - \frac{1}{2}\hat{s} \mu_{e} \oint d \Omega_{\alpha\beta}\widetilde{W}^{\alpha\beta},
\ena
where $\widetilde{W}^{\alpha\beta} = \partial^{\alpha}\widetilde{S}^{\beta} -  \partial^{\beta}\widetilde{S}^{\alpha}$ is the dual of the effective field strength. By using \eqref{duals} and \eqref{dualf} we get
\bea
\tilde{S}^{\alpha} \aeq  ( -E^{3}, B^{2}, -B^{1} ).
\ena
In the HMW configuration, the electric field $E^{3}$ vanishes and the magnetic field $B^{i}$ are constant in time. Then $S^{\alpha}(x) = ( 0, B^{2}, -B^{1} )$, and the corresponding quantum phase is
\bea
\phi^{HMW} 
\aeq \hat{s} \mu_{e} \int d \vec{r} \cdot \vec{\widetilde{S}} 
= \hat{s} \mu_{e} \oint d \vec{\Omega} \cdot \vec{ \widetilde{W} }_{B}
= - \hat{s} \mu_{e} \lambda_{m}
\ena

Similar to the AC phase, the topological and nonlocal properties of the HMW phase can be understand clearly based on this gauge symmetry. The AC and HMW effects are sensitive to only the charge distributions in the region enclosed by the particle orbit. In this sense, the AC and HMW effects are gauge invariant under the effective $U(1)$ gauge transformations: variation of the charge distributions outside of this region can not affect the AC and HMW phases. This remarkable property have to be preserved on the noncommutative space-time, otherwise we can not interpret the physical observables uniquely because of the gauge dependence. The Seiberg-Witten (SW) map \cite{SW:1999} can establish a correspondence between the quantities on noncommutative space time and the ones on commutative space time, and has been extensively studied \cite{Martin:2012aw, Brace:2001, Barnich:2001, Picariello:2002, Banerjee:2002,Fidanza:2002,Jackiw:2002}. In next section we will apply the Seiberg-Witten map for the effective $U_{jeu}(1)$ gauge symmetry, then we study the noncommutative corrections on the AC and HMW phase effects based on this method.

\section{Seiberg-Witten for neutral spinor and quantum phase effects}\label{NCACHMW}
In last section we have introduced the effective $U_{neu}(1)$ gauge symmetry for the neutral particle with anomalous dipole moments. This remarkable property have to be preserved even when the noncommutativety of space-time comes into play. The Seiberg-Witten (SW) map \cite{SW:1999} can be used to establishes a correspondence between noncommutative quantities and commutative ones. Furthermore, the SW map preserves the $U_{neu}(1)$ gauge symmetry, therefore the noncommutative corrections on the AC and HMW phases are also gauge invariant. Extensive studies \cite{Martin:2012aw, Brace:2001,Barnich:2001,Picariello:2002,Banerjee:2002,Fidanza:2002,Jackiw:2002,SubirGhosh:2005} have shown that the SW map is useful to study the noncommutative effects.
Particularly, in the exact 2+1 dimensions, the Chern-Simon term can appear, and can be relevant  when the photon fields are dynamical\cite{SubirGhosh:2005}. However, in the AC and HMW phase effects, the photon fields are static, therefore we neglect this part of effects. In this section we introduce the SW map for neutral particle with electromagnetic dipole moments, then we study the noncommutative corrections on the AC and HMW phases.

\subsection{Seiberg-Witten map}

In this subsection we apply the Seiberg-Witten map for neutral Dirac particle with an anomalous magnetic dipole moment on the noncommutative plane. Similar SW map is employed in the same way for the neutral Dirac particle with an anomalous electric dipole moment. 
On the noncommutative plane, the ordinary product is replayed by the $\star$ product defined in \eq{eq:starproduct}. The noncommutative extension of the corresponding Lagrangian \eqref{miniflag} is then, 
\bea\label{eq:nclag}
\mathcal{L}_{NC} \aeq \bar{\psi}(x) ( \pslash - m ) \star \psi(x) - \hat{s} \mu_{m} \bar{\psi}(x) \star \Sslash(x ) \star \psi(x) \,.
\ena
The corresponding effective gauge transformations are also deformed,
\bea
\label{gauge-transformation-matter}
\psi'(x) &=& U(x)\star \psi(x) ,\\
\label{gauge-transformation-gauge}
S'_{\mu}(x) &=& U(x)\star \bigg( S_{\mu}(x) -\frac{ 1 }{ \hat{s}\mu_{m} } p_{\mu} \bigg) \star U^{-1}(x).
\ena
In analogy with the SW map of the electromagnetic field theory for charged particles on noncommutative space, we apply the SW map for the neutral Dirac field $\psi$ and the effective$U_{neu}(1)$ gauge potential as follows, 
\bea\label{swmattermm}
\psi &\to & \psi - \frac{1}{2} \hat{s}\mu_{m} \theta^{\alpha\beta} S_{\alpha} \partial_{\beta}\psi \,, \\
\label{swgaugemm}
S_{\mu} &\to & S_{\mu} - \frac{1}{2} \hat{s}\mu_{m} \theta^{\alpha\beta} S_{\alpha} ( \partial_{\beta} S_{\mu} + W_{\beta\mu} )\,,
\ena
where $W_{\mu\nu} = \partial_{\mu} S_{\nu} -  \partial_{\nu} S_{\mu} $ is the effective field strength tensor. The noncommutative Lagrangian can be obtained by inserting the SW maps \eq{swmattermm} and \eq{swgaugemm} into \eq{eq:nclag}. The corresponding kinematical and interaction parts become
\bea
\mathcal{L}_{NC-K} &=& \bigg( 1 - \frac{1}{4}\hat{s}\mu_{m} \theta^{\alpha\beta} W_{\alpha\beta} \bigg) \bar{\psi}(x) ( \pslash - m ) \psi(x)\,,
\\
\mathcal{L}_{NC-I} &=& - \hat{s} \mu_{m}\bigg( 1 - \frac{1}{4}\hat{s}\mu_{m} \theta^{\alpha\beta} W_{\alpha\beta} \bigg)  \bar{\psi}(x) \star \Sslash(x ) \star \psi(x)  +
\frac{i}{2} \hat{s}\mu_{m} \theta^{\alpha\beta} \bar{\psi}(x) \gamma^{\mu} W_{\mu\alpha} \mathcal{D}_{\beta} \psi(x) \,,
\ena
where total derivatives have been drooped. Then the total Lagrangian can be written as,
\bea\label{ac-nc}
\mathcal{L}_{NC} \aeq
\bigg( 1 - \frac{1}{4}\hat{s}\mu_{m} \theta^{\alpha\beta} W_{\alpha\beta} \bigg) \bar{\psi}(x) ( i\gamma_{\mu} \mathcal{D}^{\mu}  - m )  \psi(x) +
\frac{i}{2} \hat{s}\mu_{m} \theta^{\alpha\beta} \bar{\psi}(x) \gamma^{\mu} W_{\mu\alpha} \mathcal{D}_{\beta} \psi(x) .
\ena
Because the effective covariant derivative $\mathcal{D}_{\beta}$ for the $U_{neu}(1)$ gauge symmetry is gauge invariant under the $U_{cha}(1)$ symmetry transformations, and the effective field strength tenser $W_{\mu\nu}$ is gauge invariant under both the $U_{cha}(1)$ and $U_{neu}(1)$ gauge symmetries, hence this Lagrangian is also gauge invariant under both gauge transformations. Therefore the noncommutative corrections on the AC and HMW phases can be defined unambiguously. 

The Lagrangian (\ref{ac-nc}) receives two kinds of corrections. The first term comes from the coupling between the background fields $\theta_{\mu\nu}$ and the effective field strength tensor $W_{\mu\nu}$. By our assumption this term is negligible because of vanishing field strength $W_{\mu\nu}(x)$ at the location of the matter fields $\psi(x)$. So the main effects come form the second term which depends on the covariant derivatives of the matter field $\mathcal{D}_{\beta} \psi(x)$. We will study the effects of this term on the AC and HMW phase effects in next subsection. In the rest of this subsection, let us consider the SW map for neutral Dirac particle with an anomalous electric dipole moment. Similar to the case of magnetic dipole moment, the noncommutative extension of the Lagrangian for the neutral Dirac particle with an anomalous electric dipole moment is
\bea\label{hmw}
\tilde{\mathcal{L}}_{NC} \aeq \bar{\psi}(x) ( \pslash - m ) \star \psi(x) + \hat{s} \mu_{e} \bar{\psi}(x) \star \tilde{\Sslash}(x ) \star \psi(x)~,
\ena
and the corresponding effective $U_{jeu}(1)$ gauge transformations are
\bea\label{gauge-transformation-gauge}
\psi' &=& U(x)\star \psi ~,\\
\widetilde{S}'_{\mu} &=& U(x)\star \bigg( \widetilde{S}_{\mu} +\frac{ 1 }{ \hat{s}\mu_{e} } p_{\mu} \bigg) \star U^{-1}(x)~.
\ena

Similarly, we apply the SW map for neutral particle with an anomalous electric dipole moment,
\bea
\label{dualswmattermm}
\psi &\to & \psi + \frac{1}{2} \hat{s}\mu_{e} \theta^{\alpha\beta} \widetilde{S}_{\alpha} \partial_{\beta}\psi \,, \\
\label{dualswgaugemm}
\widetilde{S}_{\mu} &\to & \widetilde{S}_{\mu} + \frac{1}{2}\hat{s}\mu_{e} \theta^{\alpha\beta} \widetilde{S}_{\alpha} ( \partial_{\beta} \widetilde{S}_{\mu} + \widetilde{W}_{\beta\mu} )\,,
\ena
where $\widetilde{W}_{\mu\nu} = \partial_{\mu} \widetilde{S}_{\nu} -  \partial_{\nu} \widetilde{S}_{\mu} $ is the dual effective field strength tensor. By using the SW maps \eqref{dualswmattermm} and \eqref{dualswgaugemm}, the Lagrangian \eqref{hmw} can be written in terms of the ordinary fields as follows,
\bea\label{hmwlag-nc}
\widetilde{\mathcal{L}}_{NC} =
\bigg( 1 + \frac{1}{4}\hat{s}\mu_{e} \theta^{\alpha\beta} \widetilde{W}_{\alpha\beta} \bigg) \bar{\psi}(x) ( i\gamma_{\mu} \widetilde{\mathcal{D}}^{\mu}  - m )  \psi(x) -
\frac{i}{2} \hat{s}\mu_{e} \theta^{\alpha\beta} \bar{\psi}(x) \gamma^{\mu} \widetilde{W}_{\mu\alpha} \widetilde{\mathcal{D}}_{\beta} \psi(x) .
\ena 
Again, this Lagrangian is gauge invariant under both the $U_{cha}(1)$ gauge transformations for charged particle and the $U_{neu}(1)$ gauge transformations for neutral particle with an anomalous electric dipole moment. There are two kinds of corrections. The correction proportional to $\big( 1 + \hat{s}\mu_{e} \theta^{\alpha\beta} \widetilde{W}_{\alpha\beta}/4 \big) $ is negligible because it is independent of the covariant derivative and vanishes at the location of the matter field $\psi(x)$. It is the second term, which depends on the dual covariant derivative of the matter field, $\widetilde{\mathcal{D}}_{\beta} \psi(x) $, results in the major corrections on the HMW phase.

\subsection{Noncommutative corrections on the Aharonov-Casher effect}
In last subsection we have obtained the noncommutative corrections on the electromagnetic dynamics of neutral Dirac particle with anomalous electromagnetic dipole moments. The major corrections come form the coupling between the background field $\theta_{\mu\nu}$, field strength $W_{\mu\nu}$ ($\widetilde{W}_{\mu\nu}$) and the (dual) covariant derivatives of the matter field $\mathcal{D}_{\beta} \psi(x)$ ($\widetilde{\mathcal{D}}_{\beta} \psi(x)$). Because in the configuration of the AC effect, the field strength $W_{\mu\nu}$ vanishes at the location of the matter particle, therefore the Lagrangian \eqref{ac-nc} can be written in a more compact form as follows,
\bea\label{ncacfinal}
\mathcal{L}_{NC} &=& \bar{\psi}(x) ( i\gamma_{\mu} \mathcal{D}_{NC}^{\mu}  - m ) \psi(x) , \\
\label{ncacfinald}
\mathcal{D}_{NC}^{\mu} &=& \bigg( g^{\mu}_{~\beta} + \frac{1}{2} \hat{s} \mu_{m} W^{\mu\alpha}\theta_{\alpha\beta} \bigg) \mathcal{D}^{\beta}.
\ena
This result is different from the previous results \cite{Mirza:2004, Li:2007,Mirza:2006} which were obtained by just using the Bopp's shift. If we use the Bopp's shift, then the effective covariant derivative on noncomutative plane is $\mathcal{D}_{NC}^{\mu} =  \big( g^{\mu\beta} + \frac{i}{2} \hat{s} \mu_{m} \theta^{\alpha\beta} \partial_{\alpha} S^{\mu} (x) \big)\mathcal{D}_{\beta}$, which is obviously not gauge invariant under the $U_{neu}(1)$ gauge transformation. Therefore noncommutative corrections can not be defined unambiguously. Remarkably, the corrections \eqref{ncacfinald} are proportional to the covariant derivative, and depend only on the effective field strength tenser $W^{\mu\alpha}$. Therefore it is invariant under the $U_{neu}(1)$ gauge transformation \eqref{swmattermm} and \eqref{swgaugemm}. In terms of the external electromagnetic fields, the correction can be written as
\bea
C^{\mu}_{~\beta}
 \equiv  \frac{1}{2} \hat{s} \mu_{m} W^{\mu\alpha}\theta_{\alpha\beta}
 \aeq
\left(\begin{array}{ccc}
0 & 0 & 0 \\
0 & \hat{s} \mu_{m}\rho_{e}  \theta_{z}/2 & 0 \\
0 & 0 & \hat{s}\mu_{m}\rho_{e} \theta_{z}/2
\end{array}\right),
\ena
where $\rho_{e} = \vec{\nabla}\cdot\vec{E}$ is the charge density of external electric field, and $\theta_{z} = \theta^{12}$. From Lagrangian \eqref{ncacfinal} one can get the equation of motion for the matter particle,
\bea\label{acmotion}
( i\gamma_{\mu} \mathcal{D}_{NC}^{\mu}  - m ) \psi(x) \aeq 0\,.
\ena
If $\psi(x)$ is a solution of the free equation, then 
\bea
\psi'(x) \aeq \exp\bigg\{ i \int \big( \hat{s} \mu_{m} S^{\mu} + i C^{\mu\nu} \mathcal{D}_{\nu} \big) d x_{\mu} \bigg\} \psi(x)
\ena
is a solution of  \eq{acmotion}. Therefore the AC phase on noncommutative plane is, 
\bea
\phi_{NC}^{AC} \aeq \phi^{AC} + \phi_{\theta-v}^{AC} + \phi_{\theta-g}^{AC}~.
\ena
The first term is the ordinary AC phase. The second term is a correction depending on the momentum of the matter particle,
\bea
\phi_{\theta-v}^{AC} 
\aeq  \int C^{\mu\nu} p_{\nu} d x_{\mu}
=  \frac{ \hat{s}}{2} \theta_{z} \mu_{m} \int  \rho_{e} \vec{p} \cdot d \vec{x} 
= \hat{s}\lambda_{e} \mu_{m} \frac{ p L \theta_{z} } { 2 A_{e} }\,,
\ena
where we have assumed the charge density $\rho_{e}$ is a constant and $\rho_{e}A_{e} = \lambda_{e}$ ($\lambda_{e}$ is the line density of electric charges) on the charged line with a cross section $A_{e}$, and vanishes otherwise; $p$ is the momentum of matter particle; $L$ is the lengthy of the closed path. The third term is a correction on the effective gauge potential
\bea
\phi_{\theta-g}^{AC} 
\aeq  - \hat{s} \mu_{m} \int C^{\mu\nu} S_{\nu} d x_{\mu}
=  - \hat{s}^{2} \mu_{m} \int  \frac{1}{2} \mu_{m} \rho_{e}^{2} \theta_{z} \vec{S} \cdot d \vec{x}
=  - \hat{s} \mu_{m} \lambda_{e} \frac{ \hat{s} \mu_{m} \lambda_{e} \theta_{z} } { 2 A_{e} }\,.
\ena
Remarkably, this correction does not depend on the momentum of matter particle. Furthermore both these two noncommutative corrections depend on the dimensionless ratio $\chi_{e}= \theta_{z}/A_{e}$. By factoring out this factor, the total correction on the AC phase can be written as
\bea
\phi^{AC}_{\theta} \aeq \phi^{AC}_{\theta-v} + \phi^{AC}_{\theta-g} = \chi_{e} K_{m}^{\hat{s}} \phi^{AC}\,,
\ena
where $K_{m}^{\hat{s}} = pL -  \hat{s} \mu_{m} \lambda_{e}$ is a dimensionless scale factor. Compared to the previous results in Refs. \cite{Mirza:2004, Li:2007,Mirza:2006}, there is an additional term proportional to $\hat{s} \mu_{m} \lambda_{e}$. Because the SW map preserves the $U_{neu}(1)$ gauge symmetry, this additional correction depends only on the electric charge distributions enclosed by the incident matter particle beam.

\subsection{Noncommutative corrections on the He-Mckellar-Wilkens  effect}
Similar to the noncommutative corrections on AC phase. The corrections come form the coupling between the background field $\theta_{\mu\nu}$ and the dual field strength $\widetilde{W}_{\mu\nu}$ vanishes in the configuration of the HMW effect. Therefore the Lagrangian \eqref{hmwlag-nc} can be written in a more compact form as follows,
\bea\label{nchmwfinal}
\widetilde{\mathcal{L}}_{NC} &=& \bar{\psi}(x) ( i\gamma_{\mu} \widetilde{\mathcal{D}}_{NC}^{\mu}  - m ) \psi(x) \,, \\
\widetilde{\mathcal{D}}_{NC}^{\mu} &=& \big( g^{\mu}_{~\beta} - \frac{1}{2} \hat{s} \mu_{e} \theta_{\alpha\beta} \widetilde{W}^{\mu\alpha} \big) \widetilde{\mathcal{D}}^{\beta}\,.
\ena
Apparently, the noncommutative corrections are invariant under both the electromagnetic $U_{cha}(1)$ gauge transformation and the effective $U_{neu}(1)$ gauge transformation. If we use the Bopp's shift method as in Refs. \cite{Wang:2007,Li:2008}, then the effective covariant derivative on noncomutative plane is $\mathcal{D}_{NC}^{\mu} =  
\mathcal{D}^{\mu} + \frac{1}{2} \hat{s} \mu_{e} \theta^{\alpha\beta} \partial_{\alpha} S^{\mu} (x) \mathcal{D}_{\beta}$, which is not invariant under the gauge transformation \eqref{dualswmattermm} and \eqref{dualswgaugemm}. Therefore the HMW phase is not well defined in the Bopp's shift approach. In terms of the external electromagnetic fields, the correction factor can be written as,
\bea
\widetilde{C}^{\mu}_{\nu} 
\aeq -\frac{1}{2}\hat{s} \mu_{e} \theta_{\alpha\beta} \widetilde{W}^{\mu\alpha} 
= 
\left(\begin{array}{ccc}
0 & 0 & 0 \\
0 & - \hat{s}\mu_{e} \rho_{m}\theta_{z}/2 & 0 \\
0 & 0 & - \hat{s}\mu_{e} \rho_{m}\theta_{z}/2 
\end{array}\right),
\ena
where $\rho_{m} = \vec{\nabla}\cdot\vec{B}$ is the magnetic monopole density of the external magnetic field. From Lagrangian \eqref{nchmwfinal} one can get the equation of motion of the matter particle,
\bea\label{hmwmotion}
( i\gamma_{\mu} \widetilde{\mathcal{D}}_{NC}^{\mu}  - m ) \psi(x) \aeq 0\,.
\ena
If $\psi(x)$ is a solution of the free equation, then 
\bea
\psi'(x) \aeq \exp\bigg\{- i \int \big( \hat{s} \mu_{e} \widetilde{S}^{\mu} - i \widetilde{C}^{\mu\nu} \widetilde{\mathcal{D}}_{\nu} \big) d x_{\mu} \bigg\} \psi(x)
\ena
is a solution of \eqref{hmwmotion}. Therefore the noncommutive HMW phase is,
\bea
\phi^{HMW}_{NC} \aeq \phi^{HMW} + \phi^{HMW}_{\theta-v} + \phi^{HMW}_{\theta-g}\,.
\ena
The first term is the original HMW phase. The second term is a noncommutative correction depending on the momentum of the matter particle,
\bea
\phi^{HMW}_{\theta-v} 
\aeq  \int \widetilde{C}^{\mu\nu} p_{\nu} d x_{\mu}
=  \frac{ \hat{s}}{2} \mu_{e} \theta_{z} \int  \rho_{m} \vec{p} \cdot d \vec{x} 
=  \hat{s}\mu_{e} \lambda_{m} \frac{ p L \theta_{z} } { 2 A_{m} } \,,
\ena
where we have assumed the magnetic charge density $\rho_{m}$ is a constant and $\rho_{m}A_{m} = \lambda_{m}$ ($\lambda_{m}$ is then the line density of magnetic charges)  on the magnetic charged line with a cross section $A_{m}$, and vanishes otherwise; $p$ is the momentum of matter particle; $L$ is the lengthy of the closed path. The third term is the noncommutative correction on the effective gauge potential,
\bea
\phi^{HMW}_{\theta-g} 
\aeq  \hat{s} \mu_{e} \int \widetilde{C}^{\mu\nu} \widetilde{S}_{\nu} d x_{\mu}
=  \hat{s} \mu_{e} \int  \frac{1}{2} \mu_{e}^{2} \rho_{m}^{2} \theta_{z} \vec{S} \cdot d \vec{x}
=  \hat{s} \mu_{e} \lambda_{m} \frac{ \hat{s} \mu_{e} \lambda_{m} \theta_{z} } { 2 A_{m} }.
\ena
Again, this correction does not depend on the momentum of matter particle. Furthermore both these two noncommutative corrections depend on the dimensionless ratio $\chi_{m}= \theta_{z}/A_{m}$. By factoring out this factor, the total correction on the HMW phase can be written as
\bea
\phi^{HMW}_{\theta} &\equiv& \phi^{HMW}_{\theta-v} + \phi^{HMW}_{\theta-g} 
  = \chi_{m} K_{e} \phi^{HMW}\,,
\ena
where $K_{e} =  - pL + \hat{s} \mu_{e} \lambda_{m}$ is a dimensionless scale factor. Compared to the previous results in Refs. \cite{Wang:2007,Li:2008}, there is an additional term proportional to $\hat{s} \mu_{e} \lambda_{m}$. Again, because the SW map preserves the $U_{neu}(1)$ gauge symmetry, this additional correction depends only on the magnetic charge distributions enclosed by the incident matter particle beam.

\section{Discussions and Conclusions}\label{conclusion}

Even through the noncommutative corrections on AC and HMW phases have also been studied in Refs.~\cite{Mirza:2004, Li:2007,Mirza:2006, Wang:2007,Li:2008}. However, the previous studies are based on the Bopp's shift method which breaks the effective $U_{neu}(1)$ gauge symmetry that is essential for the definitions of AC and HMW phase \cite{HM:1991,HM:2001,KaiMa:2012,DouglasSingleton:2015}. This results in the single velocity-dependent noncommutative correction that is hard to probe by using the ultra-cold neutron interferometer which has best precision because of the better beam coherence and interference intensity\cite{DirkDubbers:2011,ADCronin:2009}. 

The effective $U_{neu}(1)$ gauge symmetry also ensures the AC and HMW effects are affected only by the electric and magnetic charge distributions on the charged line enclosed by the particle trajectory. Charge distributions outside this region can not give any nontrivial contribution. Therefore, the effective $U_{neu}(1)$ gauge symmetry have to be preserved on the noncommutative plane for defining the AC and HMW phases unambiguously. In our approach, the SW map is employed for the effective $U_{neu}(1)$ gauge interactions on the noncommutative plane. The physical quantities on noncommutatiave and commutative space are related gauge invariantly, then we can justify if the observed new contributions are due to noncommutativity of space or not. This is important when we study the noncommutative effects experimentally.

Based on this approach, we studied the noncommutative corrections on the AC and HMW phases. Generally, the noncommutative corrections depend on the ratio between the noncommutative parameter and the cross section of charged line, $\chi_{e/m} = \theta_{z}/A_{e/m}$. This means the noncommutative corrections can be enlarged by reducing the cross section of charged line. This is one of our most important result. It provides a new scheme to improve the experimental sensitivity on the noncommutative effects. Furthermore, apart from the noncommutative corrections that depend on the particle momentum $p$ and lengthy of the trajectory $L$, there are also velocity-independent contributions. Therefore this new type of contributions are dominant in the ultra-cold matter particle interferometer. For an roughly estimation, we use the typical values in Ref.~\cite{AC:Exp}. The velocity independent term proportional to $\mu_{m}\lambda_e$ for the AC effect, and the typical value is of order 1. On the other hand, the velocity dependent term can be very large for hot neutrons, about $10^3$ for $p=0.1{\rm eV}$ and $L=1{\rm cm}$. Therefore, the velocity dependent correction is dominant for hot neutrons. However, this situation changes when we use the ultra-cold neutrons. For neutrons with velocity $1{\rm m/s}$, we have $pL \approx 10^{-3}$, which is 3 orders of magnitude smaller than the velocity independent correction. Therefore the velocity independent correction can be important when the neutron velocity smaller than $10^2{\rm m/s}$. 

In summary we provides a new approach to study the noncmmutative effects on the neutral Dirac particle with anomalous electromagnetic dipole moments. The advantages of this approach are demonstrated in the study of the noncommutative corrections on the AC and HMW effects. It is shown that this new approach can be used to define unambiguously the noncommutative corrections on the physical observables on commutative space. Therefore further applications in the studies of the noncommuative effects are promising.

\noindent\textbf{Acknowledgments}: 
K. M. is supported by the China Scholarship Council and the Hanjiang Scholar Project of Shaanxi University of Technology.  J. H. W. is supported by the National Natural Science Foundation of China under Grant No. 11147181 and the Scientific Research Project in Shaanxi Province under Grant No. 2009K01-54 and Grant No. 12JK0960. H.-X. Y. is supported in part by CNSF-10375052, the Startup Foundation of the University of Science and Technology of China and the Project of Knowledge Innovation Program (PKIP) of the Chinese Academy of Sciences.

\end{document}